\documentclass{iau}

\usepackage{amsmath}
\usepackage{graphicx}
\usepackage{multirow}

\begin{document}

\lefttitle{C. Mac Cormack et al.}
\righttitle{Prominence material observed by SoloHI beyond 0.5$\,$au}

\jnlPage{1}{7}
\jnlDoiYr{2021}
\doival{10.1017/xxxxx}

\aopheadtitle{Proceedings IAU Symposium}
\editors{C. Sterken,  J. Hearnshaw \&  D. Valls-Gabaud, eds.}

\title{Prominence material observed by SoloHI \\ beyond 0.5$\,$au}

\author{C. Mac Cormack$^{1,2}$, S. B. Shaik$^{3,4}$,  P. Hess$^{4}$,  R. Colaninno$^{4}$, T. Nieves-Chinchilla$^{1}$}
\affiliation{$^{1}$Heliospheric Physics Laboratory, Heliophysics Science Division, NASA Goddard Space Flight Center, 8800 Greenbelt Rd., Greenbelt, MD 20770, USA\\
$^{2}$The Catholic University of America, Washington, DC 20064, USA\\
$^{3}$George Mason University, Fairfax, VA 22006, USA\\
$^{4}$U.S. Naval Research Laboratory, Washington, D.C., USA
}

\begin{abstract}
With the combination of observations from in-situ and high-resolution remote sensing instruments, the Solar Orbiter (SolO) mission has become a particularly valuable mission for studying the inner heliosphere. With a field of view (FOV) of 40$^\circ$ to the east of the Sun, the Solar Orbiter Heliospheric Imager (SoloHI) is one of the six remote sensing instruments on board the Solar Orbiter (SolO) spacecraft. SoloHI's high-resolution imaging observations of the heliosphere and its higher cadence than previous generations of heliospheric imagers make it a perfect candidate to perform and complement studies of CMEs evolution through the heliosphere. In this work, we present the first prominence material detected by SoloHI along with other remote sensing instruments. We report that as the CME propagates out in the heliosphere, the associated filament material reaches a heliocentric height of $\sim$122.5$\,$R$_{\odot}$ ($\sim$0.57$\,$au). 
\end{abstract}


\maketitle

\section{Introduction}

Coronal mass ejections (CMEs) are known for being one of the main drivers of heliospheric variability. CMEs exhibit highly structured magnetic fields that play a critical role in their dynamics and evolution. These magnetic structures can vary broadly in shape, ranging from confined loop eruptions to complex helical configurations that disrupt large regions of the interplanetary medium. In these structures, we can often identify three main parts: a bright leading edge, a dark cavity, and a bright core following the cavity \citep{howard_1985,illing_1985}. The brightness of the core is mostly related to the associated prominence (or filament) material erupted from the Sun. If it is dense enough, the bright material can be observed in its evolution through the white-light coronagraphs \citep{srivastava_2000}, and becomes weaker as long as it propagates in the heliosphere. Some prominences can withstand the forces and magnetic reconfiguration associated with the eruptions and retain their structure as they propagate out in the interplanetary medium. These prominences that survive appear bright in white-light images and were observed at different heights up to 1$\,$au \citep{wood_2016,howard_2015a}, and further \citep{howard_2015b}, by the heliospheric imagers on board the Solar-Terrestrial Relations Observatory mission \citep[STEREO;][]{kaiser_2008}.

Newer solar missions have integrated high-resolution instruments that are able to fill some pre-existing gaps by providing observations closer to the Sun and out of the ecliptic plane. The Solar Orbiter Heliospheric Imager \citep[SoloHI;][]{howard_2020}, onboard the Solar Orbiter \citep[SolO;][]{muller_2020} mission, provides white-light images that allow us to track specific features detected before the CME erupts through the solar corona and the heliosphere. SoloHI is a single telescope made up of four distinct tiles with a typical field of view (FOV) of $5^{\circ}-45^{\circ}$ off the east limb of the Sun relative to the Solar Orbiter. SoloHI has a spatial resolution comparable to that of STEREO HI-1 \citep[HI-1;][]{eyles_2009}, but because the orbit of the spacecraft reaches heights to within a third of an AU, its effective resolution is significantly greater when the spacecraft is close to the Sun \citep{hess_2023}.


From the beginning of the mission until April 2023, SoloHI has detected 140 large-scale structure events, and they have been cataloged with complementary information about their whole evolution through the heliosphere. On 2022 September 23, SoloHI detected, for the first time, a prominence material that evolves in its entire FOV with the associated CME. The filament being observed through the FOV allows us to study the evolution of the CME as a whole and probe the underlying magnetic structures of the event with unprecedented image resolution beyond other available coronagraphs FOV. By combining SoloHI observations with other remote sensing data, along with the CME, we track this prominence material from its early evolution until it appears to completely diffused. 

For examining the source region, we use extreme ultraviolet (EUV) images provided by the Extreme Ultraviolet Imager \citep[EUI;][]{rochus_2020} on board SolO, the Atmospheric Imaging Assembly \citep[AIA;][]{lemen_2012} on board the Solar Dynamic Observatory \citep[SDO;][]{pesnell_2012} from Earth's point of view (POV), and the Extreme Ultraviolet Imager (EUVI) on board STEREO-A (ST-A hereafter) of the Sun-Earth Connection Coronal and Heliospheric Investigation \citep[SECCHI;][]{howard_2020} suite of instruments. To understand the coronal evolution of the prominence material, we use the observations from the Large Angle Spectroscopic Coronagraph \citep[LASCO;][]{brueckner_1995} coronagraphs (C2 and C3) on board the Solar and Heliospheric Observatory \citep[SOHO;][]{domingo_1995} mission and the COR1 and COR2 coronagraphs on board ST-A. To track the propagation of the CME through the heliosphere, we use SoloHI, HI-1, and HI-2 heliospheric imagers. 
In Section \ref{SoloHI_obs}, we describe the main features detected by SoloHI for this event. In Section \ref{source}, we discuss the source region and the CME early evolution. In Section \ref{OtherRS}, we report observations in the upper corona and heliosphere and their comparison with the observations made by SoloHI in each case. Finally, conclusions are presented in Section \ref{conclusions}.

\section{2022 September 23 event: SoloHI Observations}
\label{SoloHI_obs}

\begin{figure}[h!]
   \centering
   \includegraphics[width=0.9\hsize]{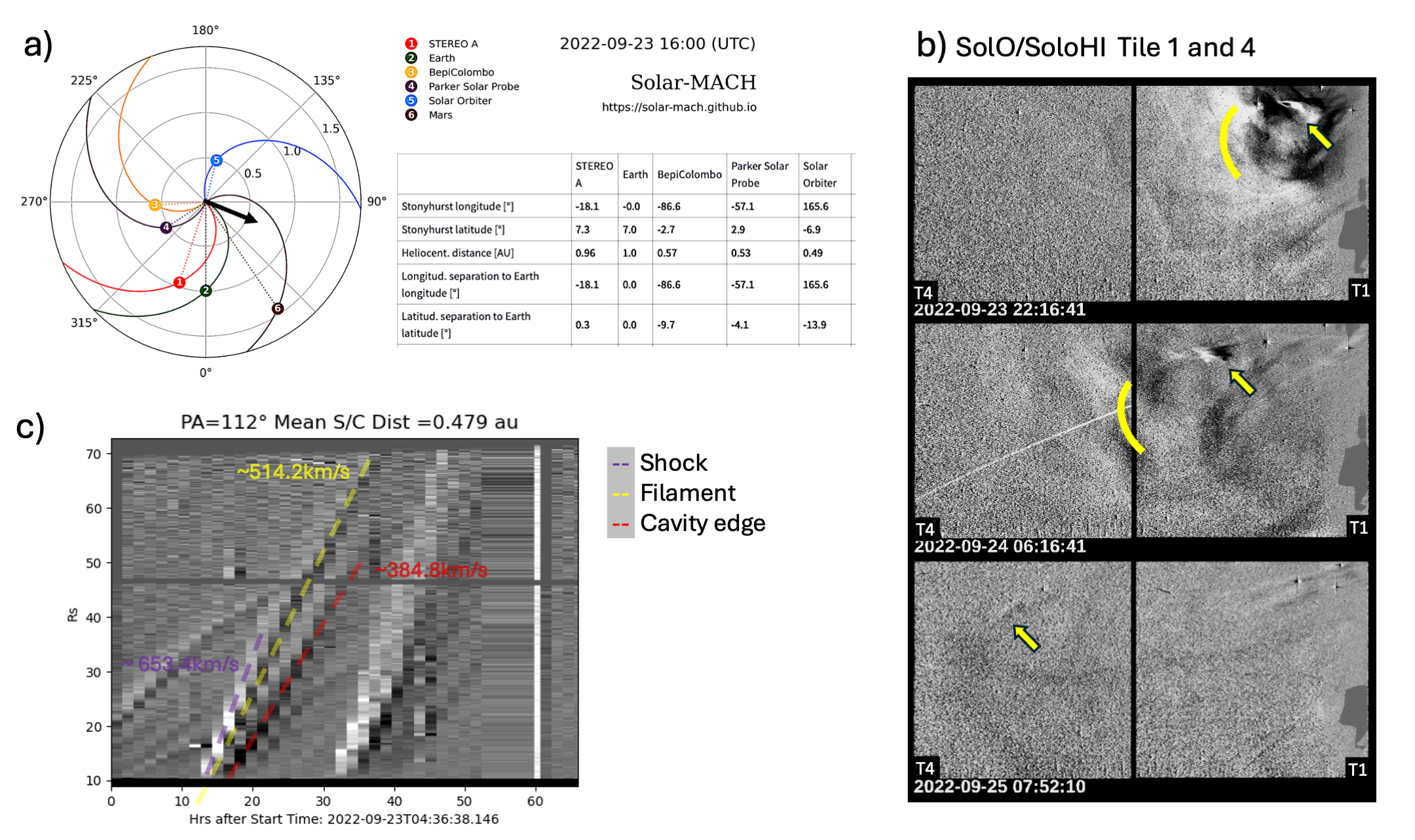}
      \caption{Panel a: Spacecraft position at the moments of the CME detection in SoloHI FOV. The black arrow indicates the longitudinal propagation of the CME. The details of the position of the spacecraft are also shown in this panel.
      Panel b: Bottom tiles of SoloHI observations on 2022 September 23 at three different times in the CME evolution. The filament material is denoted by a yellow arrow, and the cavity is indicated by the yellow line. Panel c: J-map of SoloHI observations from September 23 until September 25. Dashed violet, yellow, and red lines indicate the CME shock, filament material, and back part of the cavity, respectively. Projected velocities and estimated heights on the plane of the sky are marked.}
      \label{SoloHI-SC}
   \end{figure}

On 2022 September 23, prominence material was detected in the SoloHI FOV. Panel a of Figure \ref{SoloHI-SC} shows the spacecraft position at 16:00 UT during the detection of the CME in SoloHI. The black arrow denotes the longitudinal direction of the CME propagating at 68$^{\circ}$ of longitude. At the time of the detection of the CME in the SoloHI FOV, SolO was located at $\sim$0.49$\,$au. The first appearance of the associated CME in the SoloHI FOV was around 15:52:41$\,$UT. Mainly evolving in the ecliptic plane, the CME displays a well-defined leading edge and streamers on the sides. Although the CME evolves in the whole SoloHI FOV, the prominence material is only detected in the bottom tiles within the detector, as indicated in panel b on Figure \ref{SoloHI-SC}. A well-defined cavity can be identified below the prominence material. In the Figure, we denote the position of the filament with a yellow arrow, and a curved line indicates the low-density cavity. A movie of the CME evolution in the SoloHI FOV is attached to this article. The filament seems to vanish during its evolution in tile 4 on September 25 at 07:52:16$\,$UT. Panel c of Figure \ref{SoloHI-SC} shows a J-map made with SoloHI observations from September 23 until September 25. The evolution of the mentioned structures can be clearly identified, and a projected velocity is estimated (as indicated by the values within the panel). Although it disappears during the initial evolution of the CME in the SoloHI FOV, the shock is slightly faster than the filament. In particular, we estimated the height of the filament on the plane of the sky, being detected up to $70\,$R$_{\odot}$ or around 37$^{\circ}$ of elongation in the SoloHI FOV. Knowing the angle of propagation of the CME, we can detect the filament up to the non-projected height of $62\,$R$_{\odot}$. Multi-viewpoint analysis can provide a better description of the propagation of the filament.

\section{Source Region of the Eruption}
\label{source}

\begin{figure}[h!]
   \centering
   \includegraphics[width=0.7\hsize]{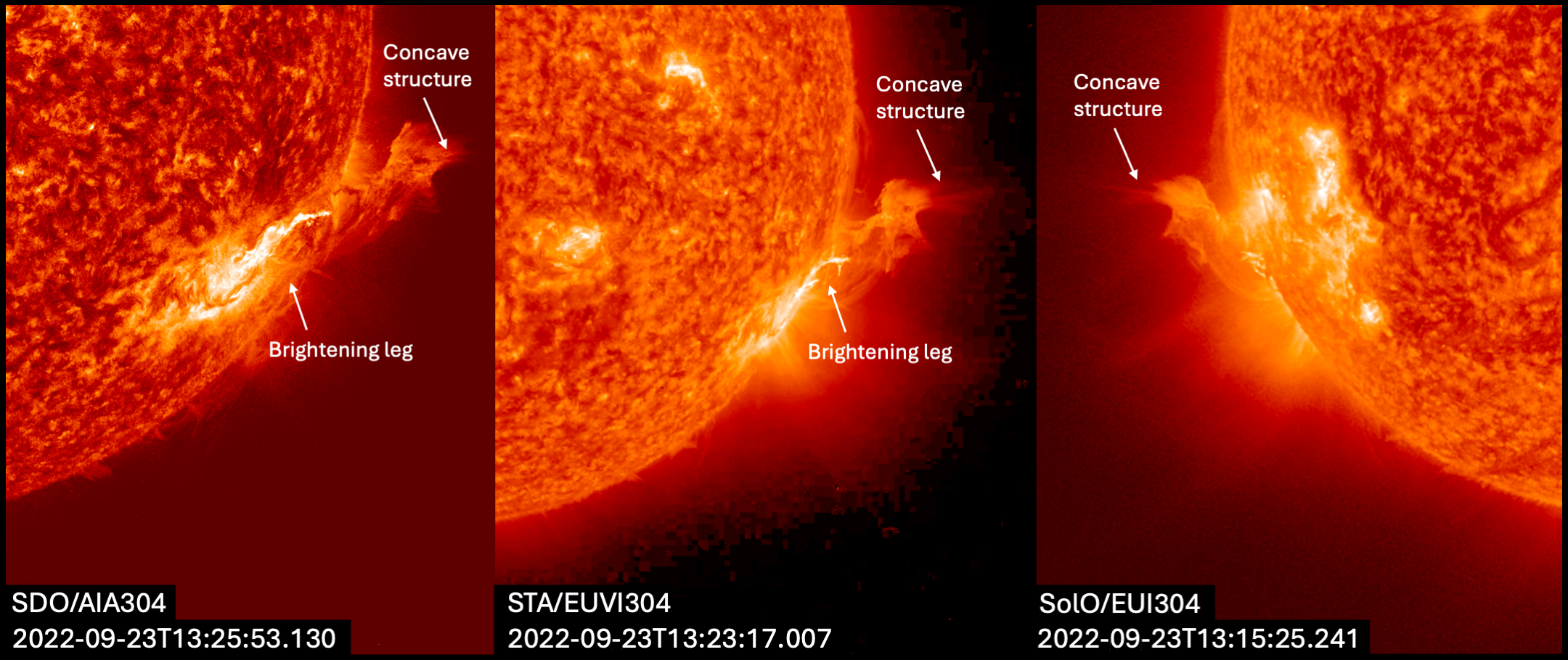}
      \caption{Observations of the eruption on 2022 September 23 by SDO/AIA 304~{\AA} (left), ST-A/EUVI 304~{\AA} (middle), and SolO/EUI 304~{\AA} (right). The erupted material can be observed in each of the three POVs. While SDO and ST-A present similar observations due to the relative similarity of their perspectives, SolO complements them on the opposite side.}
      \label{EUV_obs}
   \end{figure}

The FOVs of SDO, STEREO-A, and SolO complement one another effectively for this event. In particular, the source can be observed from Earth's POV with the SDO mission to be NOAA active region AR13102 located at south 25$^{\circ}$ and west 67$^{\circ}$. The filament starts its formation days before the eruption, but during September 22 and 23, it gets thicker and twisted, and finally erupts on 2022 September 23 at $\sim$13:25:53$\,$UT. Figure \ref{EUV_obs} shows the moment of the eruption detected from the POV of SDO/AIA, ST-A/EUVI, and SolO/EUI with the 304~{\AA} band-pass wavelength. It is easy to correlate and identify the same features in the EUV images, such as the concave structure at the highest point and the brightening leg of the filament in the AIA and EUVI images, as pointed out in the Figure.

\section{Other observations in HIs and Coronagraphs}
\label{OtherRS}

\begin{figure}[h!]
   \centering
   \includegraphics[width=0.7\hsize]{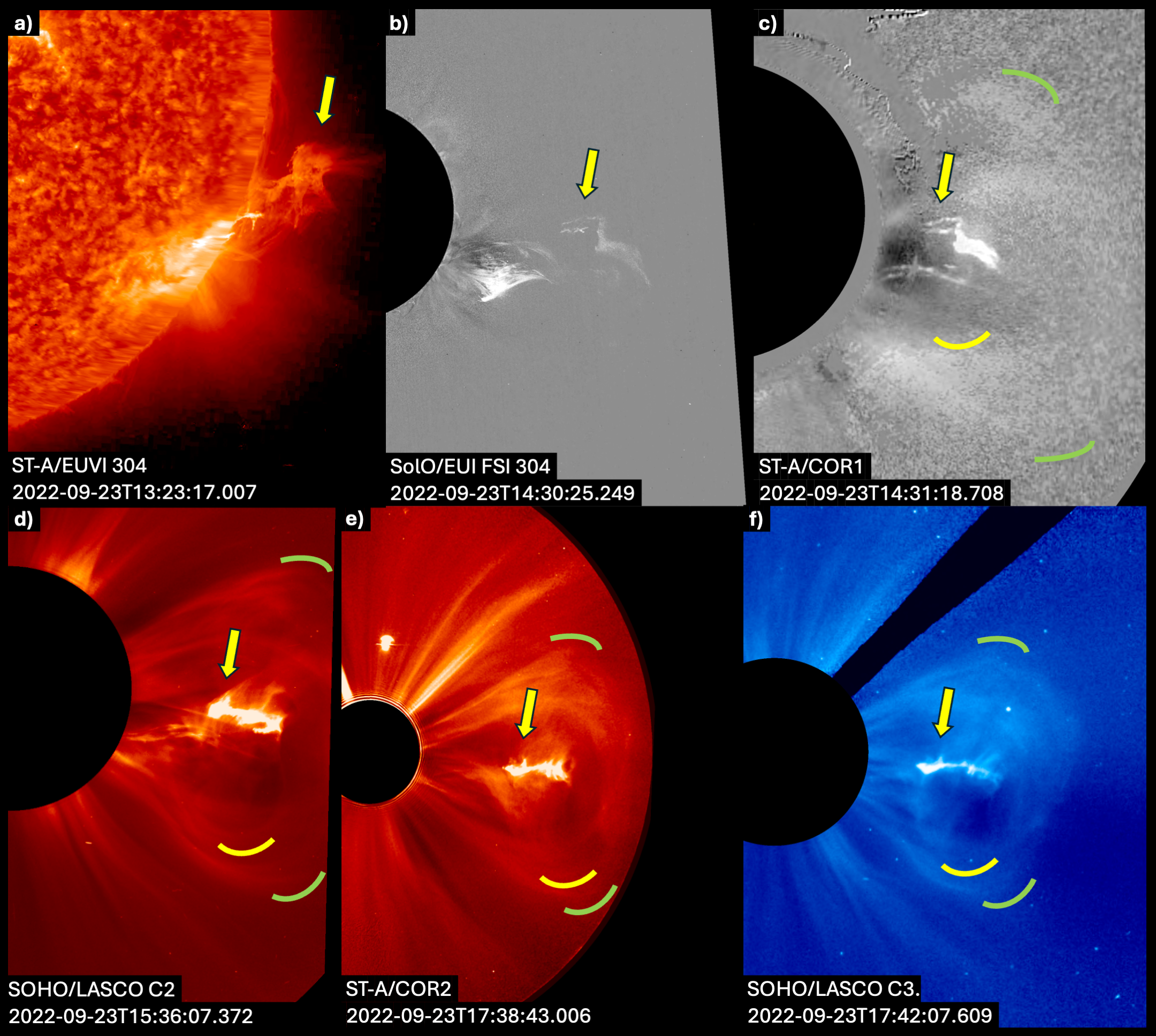}
      \caption{Evolution of the prominence material observed by different instruments on September 23, 2022. Panel a: ST-A/EUVI 304~{\AA} observations at 13:23:17$\,$UT. Panel b: Running difference image of SolO/EUI 304~{\AA} observations at 14:30:25$\,$UT. The image is flipped for an easier comparison with the other instruments' FOVs. Panel c: ST-A/COR1 observations at 14:31:18$\,$UT (also running difference image). Panel d (f): SOHO/LASCO C2 (C3) observations at 15:36:07$\,$UT (17:42:07$\,$UT). Panel e: ST-A/COR2 observations of the prominence material at 17:38:43$\,$UT. In all cases, a yellow arrow indicates the prominence material, and colored lines denote the limits of the cavity and the leading edge of the CME in the coronagraphs.}
      \label{Multi_SC}
   \end{figure}

Figure \ref{Multi_SC} shows the observations made by different remote sensing instruments on board ST-A and SOHO for the 2022 September 23 event. In all cases, the prominence material is denoted by a yellow arrow. The yellow line indicates the low-density cavity next to the prominence, and the green line indicates the leading edge of the associated CME. Panel a shows the first stage of the evolution from the FOV of ST-A/EUVI 304~{\AA} at 13:23:17$\,$UT. As we mentioned before, the eruption starts around 13:25:53$\,$UT, and the filament material can be tracked through the entire EUVI FOV (up to 1.7$\,$R$_{\odot}$). Panel b presents the eruption observed by SolO/EUI 304~{\AA} at 14:30:25$\,$UT. The image is flipped to provide a more direct comparison with the other viewpoints. With the high resolution and wide FOV of the EUI instrument, the filament can be tracked up to 3.8$^{\circ}$ at the west of the Sun (6.984$\,$R$_{\odot}$ on the plane of the sky).

EUI observations can be compared with white-light ST-A/COR1 detection shown on panel c of Figure \ref{Multi_SC} at 14:31:18$\,$UT. This is a complementary FOV between SolO and ST-A, and it can be seen that similar structures are identified with better resolution in EUI. The event appears in the COR1 FOV around 13:31:18$\,$UT, and the leading edge exits the COR1 FOV at 15:36:18$\,$UT. In the image, we detect a low-density cavity (pointed with the yellow line in the Figure) and the limits of the leading edge (green line). Panel d shows SOHO/LASCO-C2 observations at 15:36:07$\,$UT. The prominence material seems to be more elongated and travels next to the low-density cavity detected in the COR1 observations. The leading edge of the CME is also denoted by green lines. The CME appears in the LASCO-C2 FOV around 14:00:07$\,$UT, and it is no longer detected at 17:12:07$\,$UT. Panel e of Figure \ref{Multi_SC} shows ST-A/COR2 observation at 17:38:43 $\,$UT, where we can identify the same structures denoted in LASCO-C2 observations. The CME appears at 13:53:43$\,$UT in COR2 FOV, and the filament reaches the FOV limit at 21:23:43$\,$UT. In panel f, we show the SOHO/LASCO-C3 observations at 17:42:07$\,$UT. The elongated filament can be detected as long as the low-density cavity and leading edge.

\begin{figure}[h!]
   \centering
   \includegraphics[width=0.8\hsize]{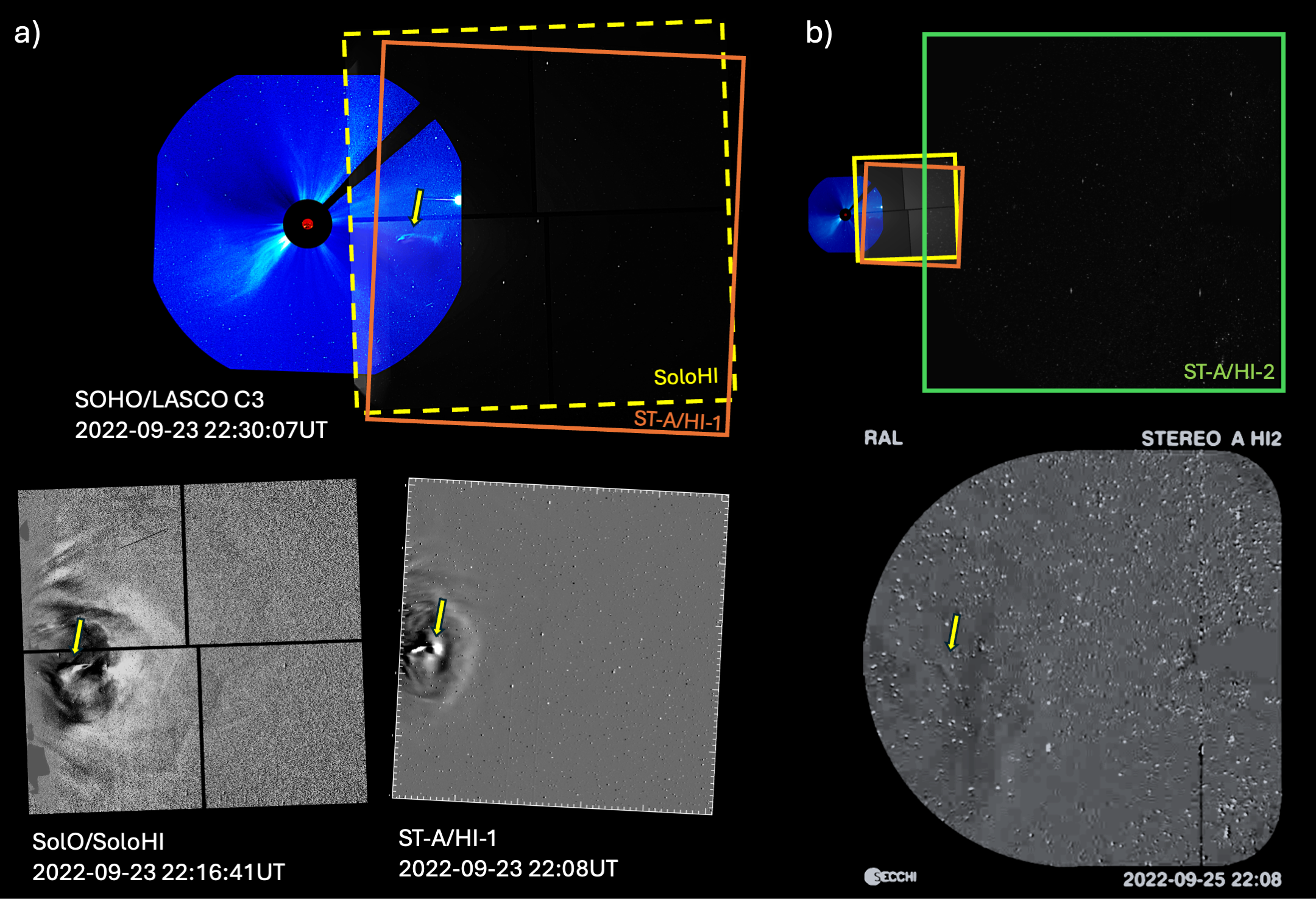}
      \caption{Panel a) Top: LASCO C3 observation of the CME on 2022 September 23 at 23:30:07$\,$UT overlaid with the complementary FOVs of SolO/SoloHI (yellow square) and ST-A/HI-1 (orange square). The dashed line indicates that the plane of the sky for SoloHI is behind ST-A/HI-1 FOV. Bottom: CME detected in SoloHI (at 22:16:41$\,$UT) and ST-A/HI-1 (at 22:08:00$\,$UT), respectively. Panel b) Top: Same as top panel a but adding ST-A/HI-2 FOV. Bottom: Filament detected by ST-A/HI-2 at 22:08:00$\,$UT on September 25, 2022.
      The yellow arrow points to the prominence material detected in all instruments.}
      \label{C3-SoloHI-STAHI1}
   \end{figure}

The LASCO-C3 FOV overlaps with SoloHI and ST-A/HI-1, as shown in Figure \ref{C3-SoloHI-STAHI1}. Panel a shows the prominence material observed by LASCO-C3 at 23:30:07$\,$UT. It still appears as an elongated filament that travels next to the low-density cavity previously identified. The yellow square represents the SoloHI FOV, and the orange square represents the ST-A/HI-1 FOV. Panel b and c represent the CME detected in SoloHI (at 22:16:41$\,$UT) and ST-A/HI-1 (at 23:08:00$\,$UT), respectively. The yellow arrow points to the prominence material detected in all instruments. In both cases we can identify the elongated prominence material in the lower limit of the FOVs. SoloHI allows us to better understand the internal structure of the CME as long as the shape of the prominence and the shock structure. 
While the ST-A/HI-1 image shows a clear front and internal structure, SoloHI reveals a clear V-shaped front and the circular flux rope still visible with the filament at the base in the first stage of the evolution. Moreover, in SoloHI movies, different structures in the elongated filament can still be identified, while the ST-A/HI-1 images are more blurred. In \citet{hess_2023}, the authors perform a comparison between the observations made by SoloHI and ST-A/HI-1, highlighting the new structures that SoloHI can describe thanks to its improved resolution.

The event is also observed by ST-A/HI-2. The CME appears in HI-2 FOV on 2022 September 24 at 14:00$\,$UT and becomes diffuse on September 25 at around 22:08$\,$UT, as shown in Panel b on Figure \ref{C3-SoloHI-STAHI1}. The structure seems to reach 35$^{\circ}$ of elongation on the ST-A/HI-2 plane of the sky. Combining this information with the location of the spacecraft at that moment, we conclude that the filament reached the height of $\sim$122.5$\,$R$_{\odot}$ ($\sim$0.57$\,$au). 


\section{Conclusions}
\label{conclusions}

We presented the first prominence material detected by SoloHI. The filament was observed at different stages of its evolution by different missions that conform to the heliophysics system of observatories. We compare the observations provided by coronagraphs and heliospheric imagers along its trajectory and track it up to a heliocentric distance of $\sim$122.5$\,$R$_{\odot}$. 

With its high-resolution images and complementary FOV, SoloHI became crucial in the multi-view analysis not only for the filament tracking along the heliosphere but also in implementing models and further analyzing the associated CME. Since the loss of STEREO-B in 2014, a gap in the multi-viewpoint studies significantly limited the study of solar transients. The Solar Orbiter mission is playing an important role in filling this gap with new orbits closer to the Sun and, eventually, out of the ecliptic. 

Although in the case presented in this study there were no in-situ signals of the CME detected on Earth or Mars, the variability of the Sun can have a strong impact on its surroundings and the heliosphere in general. The consequences of these phenomena have become increasingly relevant over the years with space exploration. The development of space-based technology and the upcoming NASA human explorations make understanding of space weather hazards a crucial task. The Solar Orbiter mission, and SoloHI in particular, is contributing key information for a better understanding of the evolution of large-scale structures in the heliosphere.



\begin{thebibliography}{}

\bibitem[Brueckner \emph{et al.}(1995)]{brueckner_1995}Brueckner, G.E., Howard, R.A., Koomen, M.J., Korendyke, C.M., Michels, D.J., Moses, J.D., and, ...: 1995, {\it Solar Physics} {\bf 162}, 357. doi:10.1007/BF00733434.

\bibitem[Domingo, Fleck, and Poland(1995)]{domingo_1995}Domingo, V., Fleck, B., and Poland, A.I.: 1995, {\it Solar Physics} {\bf 162}, 1. doi:10.1007/BF00733425.

\bibitem[Eyles \emph{et al.}(2009)]{eyles_2009}Eyles, C.J., Harrison, R.A., Davis, C.J., Waltham, N.R., Shaughnessy, B.M., Mapson-Menard, H.C.A., and, ...: 2009, {\it Solar Physics} {\bf 254}, 387. doi:10.1007/s11207-008-9299-0.

\bibitem[Hess \emph{et al.}(2023)]{hess_2023}Hess, P., Colaninno, R.C., Vourlidas, A., Howard, R.A., and Stenborg, G.: 2023, {\it Astronomy and Astrophysics} {\bf 679}, A149. doi:10.1051/0004-6361/202346907.

\bibitem[Howard \emph{et al.}(2020)]{howard_2020}Howard, R.A., Vourlidas, A., Colaninno, R.C., Korendyke, C.M., Plunkett, S.P., Carter, M.T., and, ...: 2020, {\it Astronomy and Astrophysics} {\bf 642}, A13. doi:10.1051/0004-6361/201935202.

\bibitem[Howard \emph{et al.}(1985)]{howard_1985}Howard, R.A., Sheeley, N.R., Michels, D.J., and Koomen, M.J.: 1985, {\it Journal of Geophysical Research} {\bf 90}, 8173. doi:10.1029/JA090iA09p08173.

\bibitem[Howard (2015a)]{howard_2015a}Howard, T.A.: 2015, {\it The Astrophysical Journal} {\bf 806}, 175. doi:10.1088/0004-637X/806/2/175.

\bibitem[Howard (2015b)]{howard_2015b}Howard, T.A.: 2015, {\it The Astrophysical Journal} {\bf 806}, 176. doi:10.1088/0004-637X/806/2/176.

\bibitem[Howard \emph{et al.}(2008)]{howard_2008}Howard, R.A., Moses, J.D., Vourlidas, A., Newmark, J.S., Socker, D.G., Plunkett, S.P., and, ...: 2008, {\it Space Science Reviews} {\bf 136}, 67. doi:10.1007/s11214-008-9341-4.

\bibitem[Illing and Hundhausen(1985)]{illing_1985}Illing, R.M.E. and Hundhausen, A.J.: 1985, {\it Journal of Geophysical Research} {\bf 90}, 275. doi:10.1029/JA090iA01p00275.

\bibitem[Kaiser \emph{et al.}(2008)]{kaiser_2008}Kaiser, M.L., Kucera, T.A., Davila, J.M., St. Cyr, O.C., Guhathakurta, M., and Christian, E.: 2008, {\it Space Science Reviews} {\bf 136}, 5. doi:10.1007/s11214-007-9277-0.

\bibitem[Lemen \emph{et al.}(2012)]{lemen_2012}Lemen, J.R., Title, A.M., Akin, D.J., Boerner, P.F., Chou, C., Drake, J.F., and, ...: 2012, {\it Solar Physics} {\bf 275}, 17. doi:10.1007/s11207-011-9776-8.

\bibitem[M{\"u}ller \emph{et al.}(2020)]{muller_2020}M{\"u}ller, D., St. Cyr, O.C., Zouganelis, I., Gilbert, H.R., Marsden, R., Nieves-Chinchilla, T., and, ...: 2020, {\it Astronomy and Astrophysics} {\bf 642}, A1. doi:10.1051/0004-6361/202038467.

\bibitem[Pesnell, Thompson, and Chamberlin(2012)]{pesnell_2012}Pesnell, W.D., Thompson, B.J., and Chamberlin, P.C.: 2012, {\it Solar Physics} {\bf 275}, 3. doi:10.1007/s11207-011-9841-3.

\bibitem[Rochus \emph{et al.}(2020)]{rochus_2020}Rochus, P., Auch{\`e}re, F., Berghmans, D., Harra, L., Schmutz, W., Sch{\"u}hle, U., and, ...: 2020, {\it Astronomy and Astrophysics} {\bf 642}, A8. doi:10.1051/0004-6361/201936663.

\bibitem[Srivastava \emph{et al.}(2000)]{srivastava_2000}Srivastava, N., Schwenn, R., Inhester, B., Martin, S.F., and Hanaoka, Y.: 2000, {\it The Astrophysical Journal} {\bf 534}, 468. doi:10.1086/308749.

\bibitem[Wood, Howard, and Linton(2016)]{wood_2016}Wood, B.E., Howard, R.A., and Linton, M.G.: 2016, {\it The Astrophysical Journal} {\bf 816}, 67. doi:10.3847/0004-637X/816/2/67.


\end{thebibliography}
\end{document}